\documentclass[12pt,a4paper]{article}

\usepackage{graphics}
\usepackage{latexsym}
\usepackage{amsmath}
\usepackage{amssymb}

\title{What future does the universe have?}
\author{B. Hoeneisen}
\date{\small{27 March 2001 \\
        Universidad San Francisco de Quito \\
        PO Box 17-12-841, Quito, Ecuador \\
	bruce1@fnal.gov}}
\begin{document}
\maketitle

\begin{abstract}
\noindent
We discuss the future evolution of the universe in the light
of recent observations. The apparent luminosity \textit{vs} redshift
of supernovae favor an accelerating
universe. However an Einstein-de Sitter critical universe
should not be ruled out yet.\\
universe: evolution
\end{abstract}


Let us try to understand the future of the universe in the light
of recent observations.
Consider this beautiful equation:
\begin{equation}
\left[ \frac{da}{H_0 dt} \right]^2 = \Omega_m \frac{1}{a}
+ \Omega_k + \Omega_\Lambda a^2.
\label{a}
\end{equation}
It describes the evolution of a homogeneous and isotropic universe
according to General Relativity if the density of non-relativistic
matter dominates radiation
(and other sources of pressure).
The first three terms can be obtained
from Newtonian physics and are proportional to the kinetic energy,
potential energy and total energy respectively. The last term is
due to the cosmological constant $\Lambda$ of General Relativity.
The expansion parameter $a(t)$ is normalized to $a(t_0) = 1$
today. The Hubble constant is
\begin{equation}
H_0 \equiv \left[ \frac{1}{a} \frac{da}{dt} \right]_{t_0}
\equiv 100 h_0 \cdot \mathrm{km} \cdot \mathrm{s}^{-1} \mathrm{Mpc}^{-1}
= \frac{h_0}{9.778\mathrm{Gyr}}
\label{Hubble}
\end{equation}
with $h_0 = (0.71 \pm 0.07) \times {}_{0.95}^{1.15}$\cite{pdg}.
Note that (\ref{a}) at the present time is
\begin{equation}
1 = \Omega_m + \Omega_k + \Omega_\Lambda
\label{1}
\end{equation}
so the evolution of the universe on large scales is determined
by just two independent parameters, \textit{e.g.} $\Omega_m$
and $\Omega_\Lambda$. $\Omega_m$ is the present density of
non-relativistic matter (barionic and dark) in units of the
critical density $\rho_c \equiv 3 H_0^2 / 8 \pi G$.
\begin{equation}
\Omega_k \equiv -\frac{k c^2}{H_0^2 a_0^2}
\label{omega_K}
\end{equation}
where $k = +1$, $-1$ or $0$. If $k = +1$, space has the
geometry of the 3-dimensional surface of a sphere of radius
$a_0 \times a(t)$ in 4 dimensions. If $k = -1$, space has
negative curvature. If $k = 0$ then space is flat. Finally
\begin{equation}
\Omega_\Lambda \equiv \frac{\Lambda c^2}{3 H_0^2}.
\label{omega_L}
\end{equation}

We consider the four universes listed in Table \ref{comparison}:
$(\Omega_m, \Omega_\Lambda) = (1, 0)$ is the Einstein-de Sitter
critical universe with flat space that expands for ever with a
velocity approaching zero; $(1.5, 0)$ is overdense and has a space
of finite volume that expands to a maximum and then collapses
into a Big Crunch; $(0.3, 0)$ is underdense so matter can not
halt the expansion;
and $(0.3, 0.7)$ has flat space and in the
future will expand exponentially for ever due to the cosmological
constant.

The purpose of this note is to estimate
the likelihood that each of these universes corresponds
to ours, and to understand in a simple way the issues
involved. To obtain this estimate
we compare observations with predictions using the
\textquotedblleft{chi-square}" function:
\begin{equation}
\chi^2 = \sum_{i}^{} {\frac{(\mathrm{observation}_i -
\mathrm{theory}_i)^2}{\mathrm{error}_i^2}}.
\label{chi-square}
\end{equation}
This function is written for the case when the observations are
independent so that a judicious choice of observations is needed.
In the end we shall see that the main source of uncertainty of this analysis
is the estimate of the systematic errors.
The contributions to the $\chi^2$ are listed in Table
\ref{comparison}.

\begin{table}
\begin{center}
\begin{tabular}{|ccc|c|ccccc|c|c|}\hline
& & & & \multicolumn{5}{c|} {contributions to $\chi^2$} & Total & \\
$\Omega_m$ & $\Omega_\Lambda$ & $\Omega_k$ & age & G &
B & age & $\Omega_m$ & S & $\chi^2$ & CL\\
\hline
$1.0$ & $0.0$ & $0.0$ & $9.2$Gyr & $3.2$ & $0.0$ & $2.9$ & $5.4$ & $6.1$ & $18$ & $6\%$\\
$1.5$ & $0.0$ & $-0.5$ & $8.5$Gyr & $3.7$ & $13$ & $3.8$ & $16$ & $12.0$ & $49$ & $0.0\%$\\
$0.3$ & $0.0$ & $0.7$ & $11.2$Gyr & $15$ & $25$ & $1.0$ & $0.0$ & $1.2$ & $42$ & $0.0\%$\\
$0.3$ & $0.7$ & $0.0$ & $13.3$Gyr & $12$ & $0.0$ & $0.1$ & $0.0$ & $0.0$ & $12$ & $28\%$\\
\hline
\multicolumn{4}{|r|} {degrees of freedom:} & $6$ & $1$ & $1$ & $1$ & $1$ & $10$ & \\
\hline
\end{tabular}
\end{center}
\caption{Comparison of 4 universes with observations.
Density perturbations are assumed to be scale-invariant ($n = 1$):
keeping $n$ as a free parameter does not change the conclusions
significantly, see the text.
The contributions to the $\chi^2$ are from
(G) properties of galaxies (see text); (B) Boomerang and Maxima observations;
(age) age of the oldest globular clusters;
($\Omega_m$) density measurements;
and (S) supernovae. (CL) is the probability that observations are consistent
with the model.}
\label{comparison}
\end{table}

Let us begin with observations on galaxies.
We have developed a simple model of the
hierarchical formation of galaxies\cite{Hoeneisen}.
The simulations were done with $h_0 = 0.6$.
After adjusting three parameters
(the amplitude $A$ and slope $n$ of the power spectrum of
primordial density fluctuations, and $\Omega_m$)
this model is in quantitative agreement with the
following observations: the Tully-Fisher, Faber-Jackson
and Samurai relations, the Schechter distribution
(two parameters), the galaxy-galaxy correlation
(two parameters), the fluctuation in galaxy
counts, the fluctuation in the cosmic
background radiation, and the peculiar velocities of galaxies.
To compare the model with observations we have obtained
a $\chi^2$ with 7 terms. For each $\Omega_m$ we have
minimized the $\chi^2$ by varying 2 parameters: $A$
and $n$\cite{Hoeneisen}. Therefore
for each $\Omega_m$ we are left with 5 degrees of
freedom.
The corresponding $\chi^2$'s for the four universes are
$3.2$, $3.7$, $7.6$ and $9.2$, and the
spectral indices $n$ are $1.0$, $0.9$, $1.2$ and $1.2$,
respectively.
We note that observations on
galaxies favor a universe with $\Omega_m \approx 1$,
$\Omega_\Lambda \approx 0$ (the Einstein-de Sitter universe)
and $n \approx 1$ (scale invariant spectrum)\cite{Hoeneisen}
which is quite remarkable!
In fact, for $\Omega_\Lambda = 0$, we obtain agreement
with observations provided $\Omega_m > 0.25$ and
$1.1 - 0.3\Omega_m < n < 1.4 - 0.2\Omega_m$ with $95\%$
confidence (assuming the model is correct and the
simulation volume $(92\mathrm{Mpc})^3$ is sufficiently
large)\cite{Hoeneisen}.
So all four universes considered in Table \ref{comparison}
are compatible with these observations on galaxies.
However, if we restrict the slope $n$ to $1$ as suggested by the
Boomerang and Maxima observations discussed below,
then the $\chi^2$'s for $6$ degrees of freedom
are as shown in Table \ref{comparison} column G.
Note that the two low density universes are disfavored
by observations on galaxies if we set $n = 1$:
the velocities of circular orbits of $L^*$ galaxies are too
low, the fluctuations in galaxy counts are too low
(unless $n$ is tilted to $1.25$), and the peculiar velocities
of galaxies are too low (unless $n$ is tilted to $1.3$)\cite{Hoeneisen}.

Let us turn to the Boomerang and Maxima balloon-borne experiments. By
observing the first acoustic peak of the fluctuations in the cosmic
background radiation, the Boomerang collaboration obtains
$\Omega_m + \Omega_\Lambda$ with $95\%$ confidence in the
range from $(0.88 - 1.12)$ to $(0.97 - 1.35)$ depending on the
assumed priors and parametrizations\cite{Boomerang}.
We translate this into $\Omega_m + \Omega_\Lambda = 1.08 \pm 0.16$
(all errors in this article are one standard deviation or $68\%$
confidence level unless otherwise stated).
The Boomerang collaboration also mentions that
\textquotedblleft{with reasonable
priors we find}" $\Omega_m + \Omega_\Lambda = 1.07 \pm 0.06$.
These errors are statistical. We would like to add systematic
errors, but these are hard to come by in cosmology.
A full analysis is given in \cite{Lange}.
The Maxima experiment
obtains $\Omega_m + \Omega_\Lambda = 1.0_{-0.30}^{0.15}$ at
$95\%$ confidence\cite{Maxima}.
To be somewhat conservative we finally take
$\Omega_m + \Omega_\Lambda = 1.0 \pm 0.1 \pm 0.1$
where the first error is statistical and the
last term is a rather arbitrary estimate of the systematic errors.
The corresponding contributions to $\chi^2$ are listed
in Table \ref{comparison} column B. Note that the remarkable
Boomerang and Maxima observations already rule out the two
non-flat universes listed in Table \ref{comparison}.
Let us mention that the Boomerang experiment obtains
a slope of the power spectrum of primordial density perturbations
$n = 1.0 \pm 0.1$\cite{Lange} which is consistent with
scale invariance. Maxima obtains $n = 1.08 \pm 0.1$\cite{Maxima}.

Let us consider the age of the universe as inferred from
the evolution of globular clusters. The estimates\cite{pdg}
range from $18$Gyr to $11$Gyr with additional $\pm 10\%$
errors of various origins. Thus current estimates are
$14 \pm 2$Gyr, \textquotedblleft{with a possible systematic
error of similar size}"\cite{pdg}. A recent study
obtains $11.5 \pm 1.3$Gyr\cite{Chaboyer}. We take
$14 \pm 2 \pm 2$ where the first error is statistical and
the second is a rather arbitrary estimate of the systematic
error. In Table \ref{comparison} we
show the calculated ages (for the central value of $H_0$)
and their contributions to the $\chi^2$.

Now we consider measurements of the density of the
universe. A compilation of methods and
results can be found in \cite{pdg}. The
\textquotedblleft{luminosity $\times M/L$}" methods
yield $\Omega_m$ in the range $0.1$ to $0.4$. These
measurement do not include extended dark matter halos
between galaxy clusters. In our model of galaxy
formation\cite{Hoeneisen, Hoeneisen-2} the galaxies and clusters
of galaxies have halos with density
run $\propto r^{-2}$ that extend, \textit{a grosso modo},
out to the halo of the \textquotedblleft{next}" galaxy or
cluster of galaxies (once the peculiar motion is removed).
This is possible at all times, in spite of the expansion
of the universe, because
of the ongoing hierarchical formation and merging of galaxies.
For example, if clusters of galaxies have radius $R$, and
the halos of dark matter with density run $\rho \propto r^{-2}$
extend out to $r \approx 3R$, then $\Omega_m$ measured
by the \textquotedblleft{luminosity $\times M/L$}" method
is low by a factor $\approx 3$. Such extended halos are
expected\cite{Hoeneisen, Hoeneisen-2}.
Is there any evidence to rule them out?
For these
reasons I believe that the quoted measurements of $\Omega_m$
are really lower bounds. The \textquotedblleft{baryon
fraction in galaxy cluster}" method obtains
$\Omega_m$ in the range from $0.15$ to $0.35$ but
also does not include the halos of dark matter between clusters.
The \textquotedblleft{large-scale velocity flow}"
methods yield results in the range $0.1$ to $1$\cite{pdg}.
In particular the study by Idit Zehavi and A. Dekel\cite{Zehavi}
obtains $0.4 < \Omega_m < 1.1$ for $\Omega_\Lambda = 0$ or
$0.2 < \Omega_m < 0.9$ for $\Omega_\Lambda = 0.7$ at $99\%$
confidence. The lower bound assumes $h_0 = 0.75$ and $n = 1.1$, while
the upper bound assumes $h_0 = 0.55$ and $n = 0.9$.
From the evolution of the number density of x-ray clusters
it is inferred that $\Omega_m \approx 0.74$
while $\Omega_m < 0.3$ is rejected with $95\%$
confidence\cite{Blanchard}.
For these
reasons, and because the $\chi^2$ corresponding to galaxies
already includes peculiar velocities, galaxy clustering, and
velocities of circular orbits which are used to measure
$\Omega_m$, we take a conservative
$\Omega_m = 0.3 _{-0.2}^{+0.3}$. The corresponding
contributions to the $\chi^2$ are presented in
Table \ref{comparison}.

Last, but not least, we consider the Hubble diagram of
relative luminosity (magnitude) \textit{vs} red-shift of supernovae
of type 1a. This is a very interesting
measurement. The results obtained by the
\textquotedblleft{Berkeley Supernova Cosmology
Project}"\cite{Perlmutter_1, Perlmutter, Goobar, Perlmutter_2}
have been confirmed by the
\textquotedblleft{High-z Supernova Search Team}"\cite{Leibundgut,
Garnavich, Schmidt}.
At high redshift the supernovae are fainter than expected
(from the extrapolation of low redshift supernovae) yielding
$0.75\Omega_\Lambda - \Omega_m = 0.25 \pm 0.16(\mathrm{stat.})
\pm 0.09(\mathrm{identified\ syst.}) \pm
0.47(\mathrm{evolution})$\cite{Perlmutter_1, Perlmutter_2} in the
range of interest.
The corresponding contributions to the $\chi^2$ are listed
in Table \ref{comparison} column S.
For a flat universe this measurement translates to
$\Omega_m = 1 - \Omega_\Lambda =
0.28_{-0.08}^{+0.09}(\mathrm{stat.})_{-0.04}^{+0.05}
(\mathrm{identified\ syst.})
\pm 0.27(\mathrm{evolution})$\cite{Perlmutter_1, Perlmutter_2}.
The systematic error is dominated by evolution.
The errors labeled \textquotedblleft{evolution}" above
correspond to an error in magnitude of $\pm 0.2$ for high
redshift supernovae. The preliminary  estimate of this error in
\cite{Perlmutter_2}
is $\pm 0.2$ mag. The corresponding estimate in
\cite{Schmidt} is $\pm 0.17$ mag.
In comparison note that the correction for extinction is of order $1$ mag
with and error $\approx \pm 0.15$ mag.
The challenge is to reduce the error due to evolution, and the jury is
still out. Are the high-redshift supernovae $10\%$ fainter than expected
because of \textquotedblleft{gray}"
dust, or evolution of supernovae, or
evolution of the host galaxy, or gravitational lensing,
or selection effects, or increased extinction in the host
galaxy at high redshift, or is it really due to a
cosmological constant?
Or more mundane challenges that seem to be under control
(but who knows?): how does the luminosity calibration or zero
change with redshift?; how does noise subtraction or the subtraction
of the luminosity of the host galaxy depend on luminosity or
redshift?
Radio and optical gravitational lensing observations set the limit
$-1.78 < \Omega_\Lambda - \Omega_m < 0.27$ at $95\%$
confidence\cite{Helbig}. This limit corresponds to
$-0.39 < \Omega_\Lambda < 0.64$ assuming a flat
universe\cite{Helbig} which is already cutting
well into the allowed
region of the supernovae observations.
Note that the size of the estimated systematic error due
to supernovae evolution can completely change the conclusions of this
study! If the error is half the estimate given above, then the
Einstein-de Sitter universe is ruled out.

The probabilities that observations are consistent with the
model universes is shown in column CL of Table \ref{comparison}.
These probabilities are proportional to the likelihood that
the models are correct given the observations (if in advance
we have no good reason to prefer one model over the other).
We note that the two non-flat universes are excluded by several
observations. The
preferred universe has $(\Omega_m, \Omega_\Lambda) = (0.3, 0.7)$.
Note that only one of the observations (almost) rules out
the Einstein-de Sitter universe, and in cosmology it is best
to keep an open mind.

To understand the accelerating universe
$(\Omega_m, \Omega_\Lambda) = (0.3, 0.7)$ let us compare it with
the Einstein-de Sitter universe $(1, 0)$. This latter universe
has a present age $t_0 = 2/3H_0 = 9.2$Gyr; the expansion
factor is $a = \left[3H_0t/2\right]^{2/3}$; the present
distance to the horizon is $d_H(t_0) = 2c/H_0 = 3ct_0$; and
the co-moving distance to the horizon $d_H(t)/a(t)$ will
continue increasing as $\propto t^{1/3}$ so that new galaxies
will always be entering the horizon.
The universe with a cosmological constant, $(0.3, 0.7)$,
has a present age $t_0 = 0.97/H_0 = 13.3$Gyr; the expansion
factor will grow exponentially in the future as
$a \propto \exp\left[\sqrt{\Omega_\Lambda} H_0 t \right]$;
the present
distance to the horizon is $d_H(t_0) = 3.4c/H_0 = 3.3ct_0$; and
the co-moving distance to the horizon $d_H(t)/a(t)$ will
approach $4.5c/H_0$ so that after a few $1/H_0$ no new galaxies
will enter the horizon. In addition, the galaxies that
are within the horizon will become exponentially redshifted
and dimmer: their relative luminosity will decrease as
$\propto \exp\left[-3\sqrt{\Omega_\Lambda} H_0 t \right]$ so that
after a few $1/H_0$ only the galaxies of the local group,
which are gravitationally bound, will remain visible.

Finally let us calculate the time it will take to exhaust
the fuel of the universe, \textit{i.e.} mainly hydrogen (and
also the elements from helium to manganese). We assume
that hydrogen is burned at the rate we see today.
The luminosity density of the universe is
$L_B = (2.0 \pm 0.4) \times 10^8 h_0 L_{sun} Mpc^{-3}$\cite{pdg}
with $L_{sun} \approx 3.85 \times 10^{26}$W.
The energy released per proton is $\approx 8.8$MeV.
The density of hydrogen in the universe is
$\rho_H = 0.76 \rho_B \approx 0.76 \times 0.019 \rho_{c0} / h_0^2$.
From this data we obtain a time to exhaustion of hydrogen at
the present rate of consumption of $\approx 4000$Gyr, or about $300$
times the present age of the universe.

In summary, we have considered four universes.
From Table \ref{comparison} we note that one
of them is preferred by observations: the
low density universe with a cosmological constant. The conclusion
is tentative pending a full study of possible evolutionary effects
of supernovae and a positive detection of the cosmological constant
by an independent method.
The Einstein-de Sitter universe can not be ruled out yet.
If the cosmological constant is not zero the question is: why are
terms with $\Omega_m$ and $\Omega_\Lambda$ in Equation \ref{a}
of comparable magnitude today, when one dominated by
many orders of magnitude in the past, and the other will dominate
in the future? This is the \textquotedblleft{Why now?}" problem.
Another question: why is the cosmological constant so small
compared to predictions, yet (apparently) not zero?

Let us assume that the universe is indeed
$(\Omega_m, \Omega_\Lambda) = (0.3, 0.7)$.
In this case we live in a universe that is spatially flat
and so has infinite volume; distances between far away galaxies
will expand exponentially for ever; galaxies
will cease to enter the horizon; the ones within the horizon
will dim exponentially fast until only the
local group remains visible; the galaxies of the local group,
which are gravitationally bound,
will merge into a single one;
small perturbations
$\delta \equiv (\rho - \left< \rho \right>)/\left< \rho \right>$
of the density of the universe
will cease to grow and the hierarchical formation of galaxies
will come to a halt; hydrogen fuel will become
exhausted, stars will die leaving cinders (white dwarfs which
will cool and stop shining, neutron stars and black holes),
and darkness and cold will prevail for ever. No
conscience will be there to know.

\end{document}